\documentclass[aps,twocolumn,showpacs,preprintnumbers,amsmath,amssymb,amsfonts,superscriptaddress,openany,nofootinbib]{revtex4-2}
\pdfoutput=1

\usepackage[utf8]{inputenc}
\usepackage{amsfonts}
\usepackage{amsbsy}
\usepackage{url}
\usepackage{enumerate}
\usepackage[justification=raggedright]{caption}
\usepackage{hhline}
\usepackage{graphicx,color}
\usepackage{amssymb,amsmath}
\usepackage{slashed}
\usepackage{hyperref}
\usepackage{braket}
\usepackage{simplewick}
\usepackage{ascmac}
\usepackage{latexsym}
\usepackage{pifont}
\usepackage{bm}
\usepackage{comment}
\usepackage[normalem]{ulem}
\usepackage{CJKutf8}
\usepackage{cancel}
\usepackage{tikz}
\usetikzlibrary{arrows.meta, decorations.markings}

\graphicspath{{./fig}{./}}

\begin{document}

\preprint{KEK-TH-2679, DESY-24-222}

\title{
Black String in the Standard Model
}

\author{Yu~Hamada}
\affiliation{Deutsches Elektronen-Synchrotron DESY, Notkestraßesd 85, 22607 Hamburg, Germany}
\affiliation{Research and Education Center for Natural Sciences, Keio University, 4-1-1 Hiyoshi, Yokohama, Kanagawa 223-8521, Japan}

\author{Yuta~Hamada}
\affiliation{Theory Center, IPNS, High Energy Accelerator Research Organization (KEK), 1-1 Oho, Tsukuba, Ibaraki 305-0801, Japan}
\affiliation{Graduate Institute for Advanced Studies, SOKENDAI, 1-1 Oho, Tsukuba, Ibaraki 305-0801, Japan}

\author{Hayate~Kimura}
\affiliation{Graduate Institute for Advanced Studies, SOKENDAI, 1-1 Oho, Tsukuba, Ibaraki 305-0801, Japan}

\begin{abstract}
The Swampland cobordism conjecture~\cite{McNamara:2019rup} predicts various new objects in a theory with dynamical gravity. Applying this idea to the Standard Model of particle physics, a string object is predicted. We numerically constructed such an object as a black string solution. 
\end{abstract}

\maketitle


\section{Introduction}\label{sec:intro}
The Swampland cobordism conjecture~\cite{McNamara:2019rup} states that the cobordism class of quantum gravity is trivial. In particular, the conjecture implies that all the string compactifications must be null cobordant.
This leads to new non-supersymmetric branes in string theory.
For instance, the reflection 7-brane~\cite{Dierigl:2022reg} in type IIB, and various objects in heterotic string theory are predicted~\cite{Kaidi:2023tqo,Alvarez-Garcia:2024vnr,Hamada:2024cdd,Kaidi:2024cbx}.

In addition to string theory, the Swampland conjectures apply to the Standard Model(SM) as we believe that it is in the Landscape. 
A natural cobordism group associated with the SM is $\Omega_n^{\text{Spin}}$, which is the bordism group of $n$-dimensional manifolds with the Spin structure.
When $\Omega_n^{\text{Spin}}$ is nontrivial, it implies the existence of codimension $n+1$ new branes in the SM.
In this respect, it is interesting that $\Omega_1^{\text{Spin}}$ is non-trivial, $\Omega_1^{\text{Spin}}=\mathbb{Z}_2$.
The non-trivial element is generated by the circle $S^1$ with the periodic boundary condition of the fermions.
This implies that the SM has a string-like object with the $\mathbb{Z}_2$ charge flux.
In this paper, we establish the existence of the object by numerically constructing such a string-like object as a black string solution.
The black string is supported by the Casimir energy of the SM particles and is an intrinsically quantum object.
We also compute the tension and the Hawking temperature of the black string.
Notice that the black string in the SM was considered in Ref.~\cite{Arkani-Hamed:2007ryu} with a different motivation, but the explicit construction was not established there.

Let us clarify the difference between our work and the previous studies that apply the cobordism conjecture to string theory.
In contrast to such a top-down approach, our work takes a bottom-up perspective.
The absolute stability of the $\mathbb{Z}_2$-charged black string we will construct depends on the UV completion.
This is because the relevant dimension and the tangential structure of the spacetime may differ in the UV completion (denoted by $\Omega_{\mathrm{UV}}$, see Fig.~\ref{fig:UV_IR}).
For instance, if the UV completion is given by 4d theory (\textit{e.g.}, asymptotic safety scenario of the gravity~\cite{Weinberg:1980gg}), then the relevant group would be $\Omega_{\mathrm{UV}}=\Omega_1^{\text{Spin}}$, in which case the black string is stable.
On the other hand, if the SM is UV completed by ten or eleven dimensional string theory, $\Omega_{\mathrm{UV}}$ is not equal to $\Omega_1^{\text{Spin}}$ in general, and the black string can be stable only when it corresponds to a non-trivial element of $\Omega_{\mathrm{UV}}$.
In this way, the stability of the black string depends on the UV completion.
However, even in the unstable case, we expect the black string to be a metastable configuration since its decay process should involve the UV degrees of freedom (for instance, the metric of internal space changes to trivialize the defect).

\begin{figure}
\begin{tikzpicture}[every node/.style={font=\large}, x=5cm, y=2.5cm]
    \node at (-1.2, 1.5) {UV};
    \node at (-1.2, 0) {IR};
    \node at (-0.53, 1.5) {$\Omega_{\mathrm{UV}} = 0$};
    \node at (-0.5, 1.1) {$\Omega_{\mathrm{UV}} \supset \mathbb{Z}_2$};
    \node[blue] at (0, 1.5) {metastable};
    \node[blue] at (0, 1.1) {stable};
    \node[red] at (0, 0.5) {{\small UV completion}};
    \draw[->, red, thick] (-0.5, 0.15) -- (-0.5, 0.85);
    \node at (-0.5, 0) {$\Omega_1^{\mathrm{spin}} = \mathbb{Z}_2$};
\end{tikzpicture}
\caption{The absolute stability of the black string in the SM depends on the UV completion.}
\label{fig:UV_IR}
\end{figure}

This paper is organized as follows. In Sec.~\ref{sec:setup}, we review the black string supported by the Casimir energy in the SM. In Sec.~\ref{sec:construction}, we construct the black string solution numerically. 
The tension and Hawking temperature of the black string are also computed.
We conclude in Sec.~\ref{sec:discussion}.
\section{Setup}\label{sec:setup}
In this section, we introduce the Casimir energy as a key ingredient in our study.
\subsection{Vacuum solutions}
As in Sec.~\ref{sec:intro}, we look for the black string solution.
However, as is well known, there is no event horizon whose topology is other than spherical in $d=4$ General Relativity under the assumption of the dominant energy condition (see Sec.~9.3 in \cite{hawking1975large}).
This can be seen by using general vacuum solutions of the Einstein equation with cylindrical symmetry:~\cite{Bronnikov_2020}
\begin{align}
    ds^2 =& 
    -\alpha(\Sigma z)^{\frac{4\sigma}{\Sigma}} dt^2 + \beta(\Sigma z)^{\frac{-4\sigma(1-2\sigma)}{\Sigma}} dx^2 \nonumber \\
    &\hspace{1.5em} + dz^2 + \dfrac{1}{a}(\Sigma z)^{\frac{2(1-2\sigma)}{\Sigma}} d\phi^2,    \label{eq:cylindrical_vacuum_solution}
\end{align}
where $\Sigma = \, 4\sigma^2-2\sigma+1$, $\sigma$ takes real arbitrary value,\footnote{Especially, $\sigma=0$ corresponds to the flat metric.} and $z$, $x$, and $\phi$ parametrize the radial direction, the longitudinal direction and the angle around the $x$ axes, respectively (see Fig.~\ref{fig:coordinate}).
The coefficients $\alpha, \beta$ are arbitrary positive numbers corresponding to rescaling of $t, x$.

The curvature singularity is located at $z=0$, and there is no horizon that hides the singularity.
In order to have a black string solution without naked singularity, we consider the quantum effect.

\subsection{Black string horizon supported by quantum effect}
\begin{figure}
    \centering
    \includegraphics[width=0.5\linewidth]{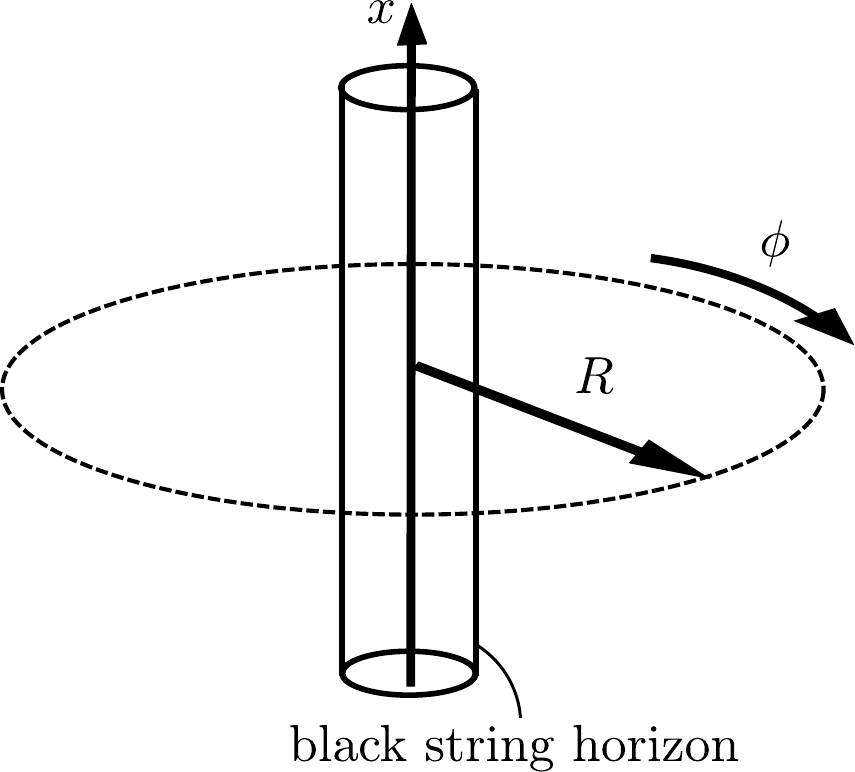}
    \caption{Our coordinate system.}
    \label{fig:coordinate}
\end{figure}

The key observation to construct the black string is that the fields going around the string feel the non-trivial boundary condition~\cite{Arkani-Hamed:2007ryu}.
Consequently, the Casimir energy is generated as a quantum effect.
The Casimir energy associated to the $S^1$ surrounding the string is (see e.g.~\cite{Hamada:2017yji})
\begin{align}
\begin{split}
    V_{\mathrm{Cas}} = -\sum_{\text{particle}}&(-1)^{2s_p}n_p \frac{m_p^4}{2\pi^2} \\
    &\times\sum_{n=1}^\infty \frac{\cos(2\pi n\theta_p)}{(2\pi R m_p n)^2}K_2(2\pi R m_p n). \label{eq:casimir_energy}
\end{split}
\end{align}
The sum is over particles in the SM, $s_p$ is its spin, $m_p$ is its mass, and $n_p$ is the real degrees of freedom. 
$R$ is the radial distance from the string, which is along with the vertical axis $x$ of the cylindrical coordinate (see Fig.~\ref{fig:coordinate}). 
The parameter $\theta_p$ corresponds to the boundary condition when the particle moves around the string. 
The periodic and anti-periodic boundary conditions correspond to $\theta_p=0$ and $\theta_p=1/2$, respectively.
We consider the periodic boundary condition both for bosons and fermions.\footnote{This corresponds to the non-trivial element of $\Omega_1^{\text{Spin}}$.}

As the Casimir energy violates the dominant and null energy conditions, the no-go theorem of the black string is circumvented.

\section{Black String Solution}\label{sec:construction}
We look for black string solutions in the presence of the Casimir energy.
In this paper, we assume that neutrinos are Majorana and the normal hierarchy, whose lightest mass is 0 (the heaviest neutrino mass is $m_3=0.05$\,eV~\cite{ParticleDataGroup:2024cfk}).
See Supplemental Material I for the plot of the Casimir energy in this setup.
In this section, we first introduce the metric ansatz and boundary conditions, and then solve the Einstein equation numerically.
\subsection{Metric ansatz and boundary conditions}
\paragraph*{Extremal case.--}
It is natural to assume the boost invariance in the direction in which this black string extends ($x$-axis), as in the case of extremal black $p$-branes~\cite{Horowitz:1991cd}.
Therefore, we employ the following metric ansatz:
\begin{align}
    ds^2 = A(z)^2(-dt^2+dx^2) + \frac{M_P^2}{m_3^4}dz^2 + R(z)^2 d\phi^2, \label{eq:extremal_ansatz}
\end{align}
where $M_Pz/m_3^2$ is a proper distance of radial direction.
The scale $M_P/m_3^2$ is introduced in such a way that $z$ is dimensionless and the equation of motion (EOM) is simplified.
Note that $z\to+\infty$ and $z\to-\infty$ correspond to asymptotic infinity and the horizon, respectively.

As for the boundary conditions,
we require the metric to approach to the flat for $z\to\infty$ and to the $AdS_3\times S^1$ for $z\to-\infty$ (see \cite{Arkani-Hamed:2007ryu} and Supplemental Material I for details):
\begin{align}
    &A(z) \xrightarrow{z\to-\infty} e^{\frac{M_P}{m_3^2}\frac{z}{l_{\mathrm{AdS}}}}, \ \ \ \ \
    R(z) \xrightarrow{z\to-\infty} R_0^\mathrm{ext}, \label{eq:bc_extremal_horizon} \\
    &A(z) \xrightarrow{z\to\infty} \mathrm{const.}, \ \ \ \ \ R(z) \xrightarrow{z\to\infty} \mathrm{const.} \times \dfrac{M_P}{m_3^2}z. \label{eq:bc_extremal_infty}
\end{align}
Here $l_{\mathrm{AdS}}$ is the $AdS_3$ radius near the horizon, and
$R_0^\mathrm{ext}$ is the horizon size at which radion potential (28) has a minimum.
The coefficient of $z$ of $R(z)$ as $z\to\infty$ determines a deficit angle of the black string.

\smallskip
\paragraph*{Non-extremal case.--}
Motivated by the vacuum solution \eqref{eq:cylindrical_vacuum_solution}, we consider the following metric ansatz:
\begin{align}
    ds^2 = -A(z)^2 dt^2 + B(z)^2 dx^2 + \frac{M_P^2}{m_3^4}dz^2 + R(z)^2 d\phi^2. \label{eq:non_extremal_ansatz}
\end{align}
This ansatz no longer has Lorentz invariance on the $(tx)$-plane.
For $z\to\infty$, Casimir energy is suppressed, and black string solutions should approach Eq.~\eqref{eq:cylindrical_vacuum_solution}.
This provides the boundary conditions at infinity.
For instance, we see $R(z) \to (1/\sqrt{a})\cdot(\Sigma z)^{\frac{(1-2\sigma)}{4\sigma^2-2\sigma+1}}$.
The coefficient $a$ of $R(z)$ is physical and determines the deficit angle. 

On the other hand, for small $z$, we assume that the horizon is located at finite $z$.\footnote{This is again the analogy of the non-extremal RN black hole case, where its throat region has a finite length of proper distance.}
Without loss of generality, we put the horizon at $z=0$.
We impose
\begin{align}
    A(0) = 0, \ \ \  B^\prime(0) = 0, \ \ \  R(0) = R_0, \ \ \  R^\prime(0) = 0,
    \label{eq:bc_at_horizon}
\end{align}
at the horizon to avoid the curvature singularity at $z=0$ (see Supplemental Material I\hspace{-1.2pt}I. 
Here we have introduced a parameter $R_0$ to parameterize the solutions.
As varying $R_0$, one obtains a family of the solutions.

\subsection{Einstein equations}
We derive the Einstein equation from the action
\begin{align}
    S = \int d^4x \sqrt{-g} \left(\dfrac{1}{2}M_P^2\mathcal{R}-V_{\mathrm{Cas}} + \ldots\right).
\end{align}
The variation of the action by metric gives
\begin{align}
\begin{aligned}
    -&\mathcal{R}_{\mu\nu} + \dfrac{g_{\mu\nu}}{2}\mathcal{R} 
    = g_{\mu\nu}\frac{V_{\mathrm{Cas}}}{M_P^2} + \delta_{\mu3}\delta_{\nu3} \frac{R^3}{M_P^2} \dfrac{\partial V_{\mathrm{Cas}}}{\partial R}. \label{eq:Einstein_equation}
\end{aligned}
\end{align}
This is the Einstein equation to be solved. The right-hand side is the energy-momentum tensor of the Casimir energy.

\smallskip
\paragraph*{Extremal case.--}
The Einstein equation \eqref{eq:Einstein_equation} with the ansatz~\eqref{eq:extremal_ansatz} is 
\begin{align}
    &R^{\prime\prime}+R^\prime\frac{A^\prime}{A}+R\frac{A^{\prime\prime}}{A}= -R \tilde{V}_{\mathrm{Cas}},
    \label{eq:Einstein_extremal_1}\\
    &\frac{A^\prime}{A}\left(2R^\prime+R\frac{A^\prime}{A}\right) = -R \tilde{V}_{\mathrm{Cas}},
    \label{eq:Einstein_extremal_2}\\
    &\frac{A^{\prime2}}{A^2}+2\frac{A^{\prime\prime}}{A} =
    -\tilde{V}_{\mathrm{Cas}}-R\partial_R \tilde{V}_{\mathrm{Cas}},
    \label{eq:Einstein_extremal_3}
\end{align}
where the prime is the derivative with respect to $z$, and $\tilde{V}_{\mathrm{Cas}} := V_{\mathrm{Cas}}/m_3^4$.
There are three equations despite only two unknown functions $A(z)$ and $R(z)$.
This is because the first order equation~\eqref{eq:Einstein_extremal_2} is the Hamiltonian constraint.

Eq.~\eqref{eq:Einstein_extremal_2} can be solved for $A^\prime/A$:
\begin{align}
    \frac{A^\prime}{A}=
    -\frac{R^\prime}{R}+\sqrt{\frac{R^{\prime 2}}{R^2}-\tilde{V}_{\mathrm{Cas}}},
    \label{eq:extremal_eq_of_A}
\end{align}
where we have chosen the branch to be consistent with the boundary condition \eqref{eq:bc_extremal_infty}.
By using (\ref{eq:Einstein_extremal_1}, \ref{eq:Einstein_extremal_3}, \ref{eq:extremal_eq_of_A}) we obtain
\begin{align}
    &R^{\prime\prime} + \gamma R^\prime = -R\left(\tilde{V}_{\mathrm{Cas}}-\frac{1}{2}R\partial_R \tilde{V}_{\mathrm{Cas}}\right),
    \label{eq:shooting_of_R} \\
    &\gamma = -2\left(\frac{R^\prime}{R}-\sqrt{\frac{R^{\prime 2}}{R^2}-\tilde{V}_{\mathrm{Cas}}}\right).
\end{align}
Eq.~\eqref{eq:shooting_of_R} has an interpretation as a classical mechanical EOM with the friction $\gamma$ and the potential $U$:
\begin{align}
    \partial_R U = R\left(\tilde{V}_{\mathrm{Cas}}-\frac{1}{2}R\partial_R \tilde{V}_{\mathrm{Cas}}\right),
\end{align}
with $z$ being the ``time variable.''
This potential has a local maximum at the point $R=R_0^\mathrm{ext}$ where $\partial_R U=0$ is satisfied.
The solution we are looking for is the one with $\left.R\right|_{z=-\infty}=R_0^\mathrm{ext}$ and $\left.R\right|_{z=+\infty}=+\infty$.

There are three integration constants in the solution of \eqref{eq:extremal_eq_of_A} and \eqref{eq:shooting_of_R}.
One of them is fixed by the condition above.
The other two are unphysical parameters corresponding to the rescaling of $t, x$, and the shift of $z$.

\smallskip
\paragraph*{Non-extremal case.--}
With the metric ansatz~\eqref{eq:non_extremal_ansatz}, the Einstein equation~\eqref{eq:Einstein_equation} becomes
\begin{align}
    &\frac{R^{\prime\prime}}{R}+\frac{R^\prime}{R}\frac{B^\prime}{B}+\frac{B^{\prime\prime}}{B}=-\tilde{V}_{\mathrm{Cas}},
    \label{eq:Einstein_nonextremal_1}\\
    &\frac{R^{\prime\prime}}{R}+\frac{R^\prime}{R}\frac{A^\prime}{A}+\frac{A^{\prime\prime}}{A}=-\tilde{V}_{\mathrm{Cas}},
    \label{eq:Einstein_nonextremal_2}\\
    &\frac{A^\prime}{A}\frac{B^\prime}{B}+\frac{B^\prime}{B}\frac{R^\prime}{R}+\frac{R^\prime}{R}\frac{A^\prime}{A}=-\tilde{V}_{\mathrm{Cas}},
    \label{eq:Einstein_nonextremal_3}\\
    &\frac{A^\prime}{A}\frac{B^\prime}{B}+\frac{A^{\prime\prime}}{A}+\frac{B^{\prime\prime}}{B}=-\tilde{V}_{\mathrm{Cas}}-R\partial_R \tilde{V}_{\mathrm{Cas}},
    \label{eq:Einstein_nonextremal_4}
\end{align}
Among them, the first order equation~\eqref{eq:Einstein_nonextremal_3} is again a constraint equation.
Eq.~\eqref{eq:Einstein_nonextremal_3} can be solved for $A^\prime/A$:
\begin{align}
    \frac{A^\prime}{A}=\dfrac{-\tilde{V}_{\mathrm{Cas}}B R-B^\prime R^\prime}{(BR)^\prime}. \label{eq:non_extremal_eq_of_A}
\end{align}
By using (\ref{eq:Einstein_nonextremal_2}, \ref{eq:Einstein_nonextremal_4}, \ref{eq:non_extremal_eq_of_A}), we obtain
\begin{align}
    \frac{R^{\prime\prime}}{R}+\frac{
    B^{\prime2} R^\prime}{B (BR)^\prime}&+\frac{B R^\prime}{(BR)^\prime}\tilde{V}_{\mathrm{Cas}} 
    -\frac{R}{2}\partial_R \tilde{V}_{\mathrm{Cas}}=0.
    \label{eq:Einstein_nonextremal_5}
\end{align}
By solving Eqs.~\eqref{eq:Einstein_nonextremal_1} and \eqref{eq:Einstein_nonextremal_5}, we obtain $B$ and $R$.
Then, the solution for $A$ is obtained by substituting $B$ and $R$ into \eqref{eq:non_extremal_eq_of_A}.
Notice that there are five integration constants.
Four of them are fixed by the boundary condition \eqref{eq:bc_at_horizon},
leaving one unphysical parameter corresponding to the rescaling of $x$.
Thus, the only physical parameter is the horizon size $R_0$.

\subsection{Numerical Results}
\begin{figure}
    \centering
    \includegraphics[width=\linewidth]{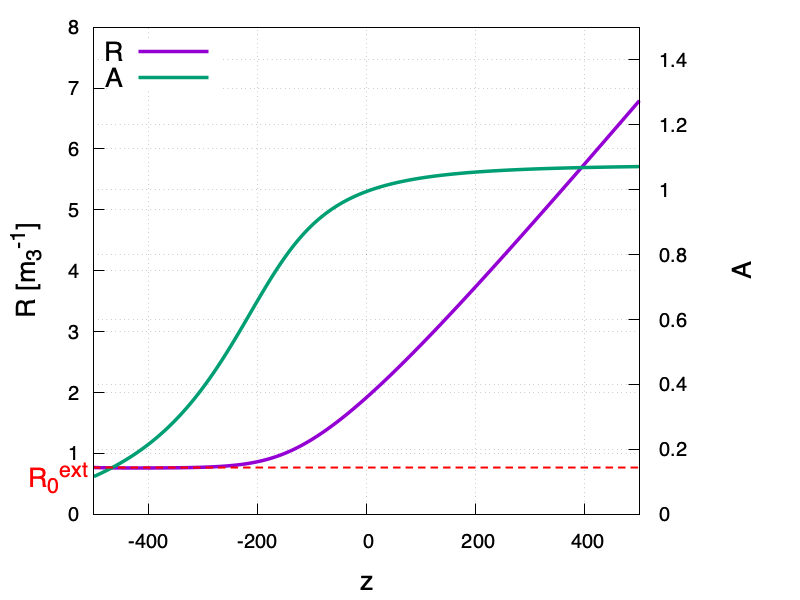}
    \caption{The extremal solution for $R$ and $A$ as a function of $z$. The left axis is for $R$, and the right axis is for $A$.}
    \label{fig:extremal_solutions}
\end{figure}

\paragraph*{Extremal case.--}
We utilize two different methods independently, the shooting method and the modified gradient flow method~\cite{Chigusa:2019wxb}
to solve \eqref{eq:extremal_eq_of_A} and \eqref{eq:shooting_of_R} with the boundary conditions (\ref{eq:bc_extremal_horizon}, \ref{eq:bc_extremal_infty}),
where the two methods show good agreement in the solutions.
The solutions are shown in Fig.~\ref{fig:extremal_solutions},
in which $R$ interpolates the $S^1$ compacted region where $R$ is almost constant and the flat region where $R$ is proportional to $z$.

The important physical parameter is $R^\prime(\infty)$. For our solution,
\begin{align}
    R^\prime(\infty) = \mathcal{O}\left(10^{-2}\right) m_3^{-1}.
\end{align}
This indicates that the opening angle is $\theta_o = 2\pi R^\prime(\infty)\times(m_3^2/M_P) \sim 2\pi \times 10^{-2} \times (m_3/M_P)\ll1$ and the deficit angle is $\theta = 2\pi - \theta_o = \mathcal{O}(1)$.

The solution for $A$ also interpolates the $AdS_3$ region where $A$ exponentially decays with $AdS$ radius $l_{\mathrm{AdS}} \sim 6.2 \times 10^{-3} \times (M_P/m_3^2)$ and the flat region where $A$ is almost constant.
Since $A$ goes to $0$ as $z\to -\infty$,
$z = -\infty$ is the coordinate singularity and corresponds to the horizon.

Let us calculate the Hawking temperature.
We start from the metric~\eqref{eq:non_extremal_ansatz} and change the radial coordinate from $z$ to $R$.
\begin{align}
    ds^2 = -A(z)^2 dt^2 + \dfrac{M_P^2}{m_3^4}\dfrac{dR^2}{R^{\prime 2}} + B(z)^2 dx^2 + R^2 d\phi^2,
\end{align}
where $R^\prime \equiv \dfrac{dR}{dz}$. It is sufficient to consider only the $(tR)$ sector of the metric. We concentrate on the metric near the horizon, and the metric components are expanded as
\begin{align}
    &A^2 \sim \tilde{\alpha} (R-R_0),
    \ \ \ \ \
    R^{\prime 2} \sim \tilde{\beta}(R-R_0), \nonumber \\
    \ \ \ \ \
    &\tilde{\alpha} := \left. \dfrac{\partial A^2}{\partial R} \right|_{\mathrm{horizon}},
    \ \ \ \ \
    \tilde{\beta} := \left. \dfrac{\partial R^{\prime 2}}{\partial R} \right|_{\mathrm{horizon}}.
\end{align}
Then, the $(tR)$ sector of the metric is
\begin{align}
    ds^2 = -\tilde{\alpha} \xi dt^2 + \dfrac{M_P^2}{m_3^4}\dfrac{d\xi^2}{\tilde{\beta} \xi} + \ldots,
    \ \ \ \ \
    \xi := R-R_0.
\end{align}
Furthermore, we introduce the new coordinate $\zeta$ as
\begin{align}
    \dfrac{M_P^2}{m_3^4}\dfrac{d\xi^2}{\tilde{\beta} \xi} = d\zeta^2,
\end{align}
and the metric becomes
\begin{align}
    ds^2 = -\dfrac{\tilde{\alpha} \tilde{\beta}}{4} \left(\dfrac{m_3^2}{M_P}\right)^2 \zeta^2 dt^2 + d\zeta^2 + \cdots.
\end{align}
When the metric is Wick rotated ($t \to -i\tau$), the metric has a conical singularity, in general. In order to avoid it, the periodicity of $\tau$ has to be
\begin{align}
    (\text{periodicity of }\tau)=\dfrac{4\pi}{\sqrt{\tilde{\alpha} \tilde{\beta}}}\left(\dfrac{M_P}{m_3^2}\right).
\end{align}
This corresponds to the inverse temperature in the context of the quantum statistical physics, and therefore the Hawking temperature is given by
\begin{align}
    &T_\mathrm{H} = \dfrac{\sqrt{\tilde{\alpha}\tilde{\beta}}}{4\pi}\left(\dfrac{m_3^2}{M_P}\right).
\end{align}
For the extremal black string, the Taylor expansion of $A(R)^2$ starts from $(R-R_0^\mathrm{ext})^2$. 
This means that $\tilde{\alpha} = 0$ and $T_H=0$, as expected.

\begin{figure}
    \centering
    \includegraphics[width=1\linewidth]{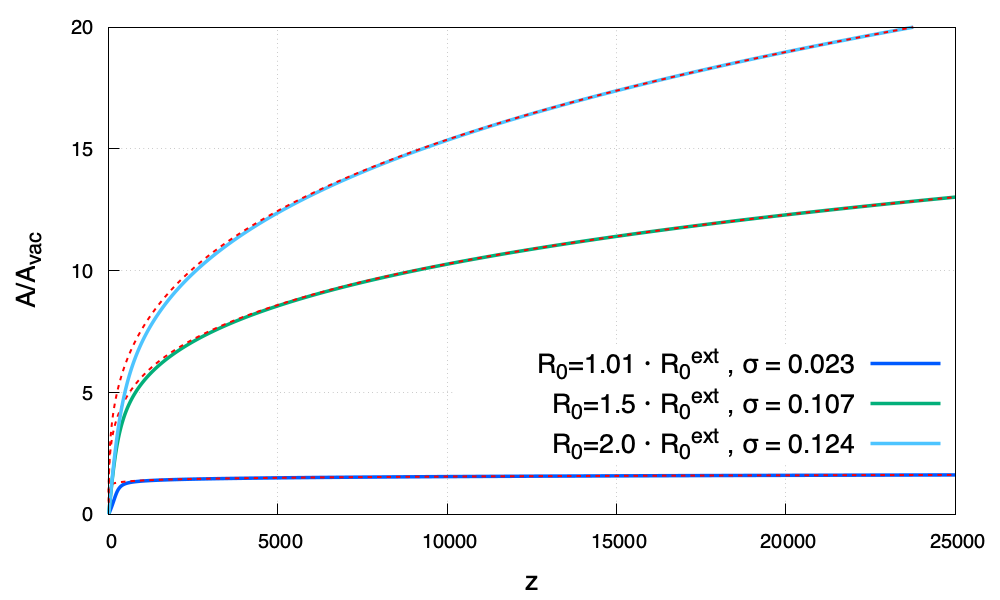}
    \caption{The non-extremal solution for $A$ as a function of $z$, where $A_\text{vac}:=(M_P/m_3^2)^\frac{2\sigma}{\Sigma}$. The normalization of $A$ is chosen in such a way that $\alpha=1$ is realized for large $z$~\eqref{eq:cylindrical_vacuum_solution}.}
    \label{fig:non_extremal_A}
\end{figure}

\begin{figure}
    \centering
    \includegraphics[width=1\linewidth]{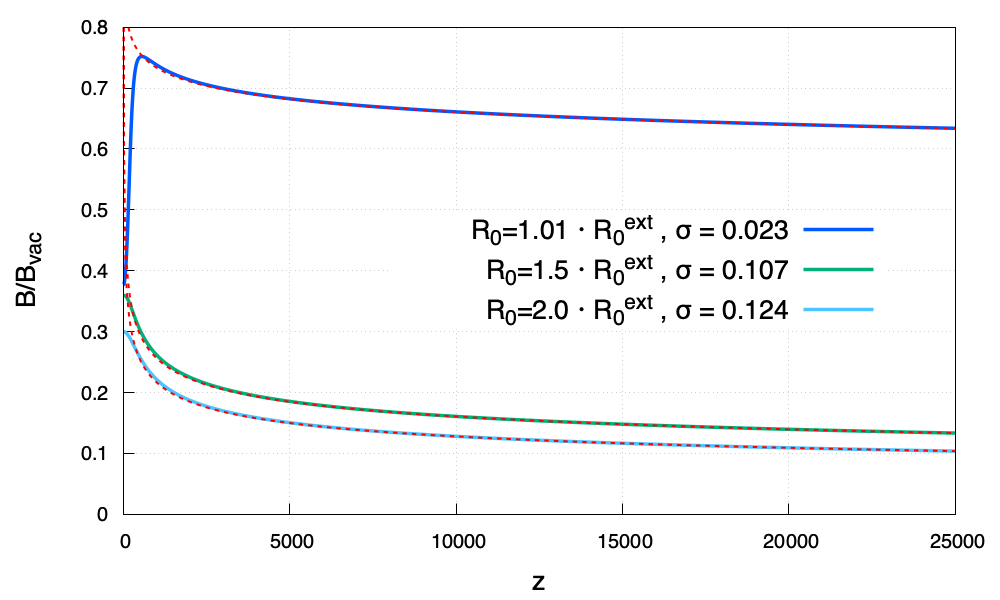}
    \caption{The non-extremal solution for $B$ as a function of $z$, where $B_\text{vac}:=(M_P/m_3^2)^\frac{-2\sigma(1-2\sigma)}{\Sigma}$. The normalization of $B$ is chosen in such a way that $\beta=1$ is realized for large $z$~\eqref{eq:cylindrical_vacuum_solution}.}
    \label{fig:non_extremal_B}
\end{figure}

\begin{figure}
    \centering
    \includegraphics[width=\linewidth]{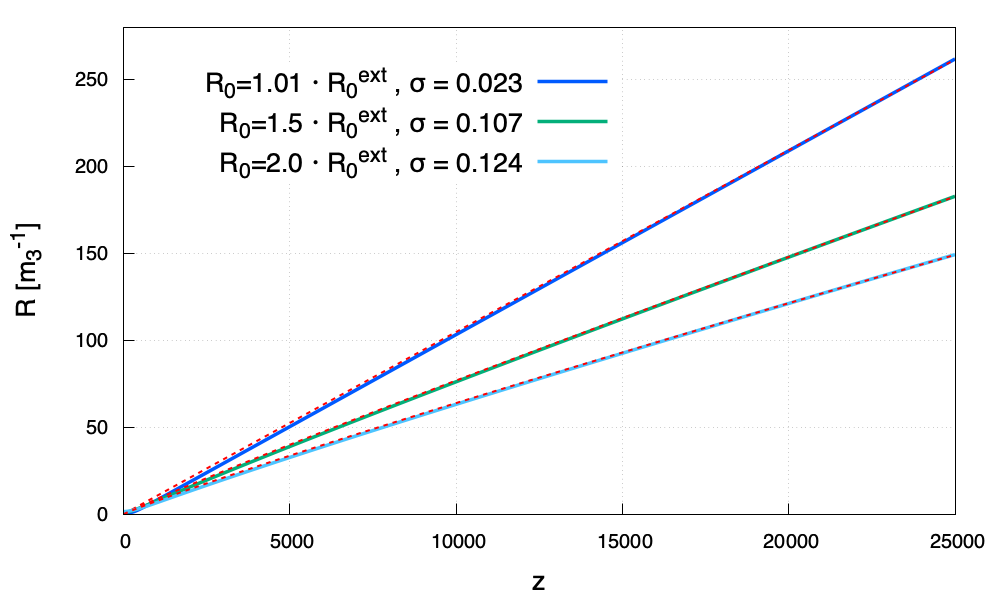}
    \caption{The non-extremal solution for $R$ as a function of $z$.}
    \label{fig:non_extremal_R}
\end{figure}

\smallskip
\paragraph*{Non-extremal case.--}
We simply integrate the differential equations \eqref{eq:Einstein_nonextremal_1}, \eqref{eq:non_extremal_eq_of_A} and \eqref{eq:Einstein_nonextremal_5} starting from the horizon $z=0$. 
The numerical solutions for several different values of the horizon size $R_0$ are plotted in Figs.~\ref{fig:non_extremal_A} to \ref{fig:non_extremal_R}.
The solutions approach the vacuum ones \eqref{eq:cylindrical_vacuum_solution}(red dashed line) for large $z$, as it should be.
Here, $\sigma$ and $a$ are determined by fitting to the numerical solutions.

\begin{figure}
    \centering
    \includegraphics[width=1.05\linewidth]{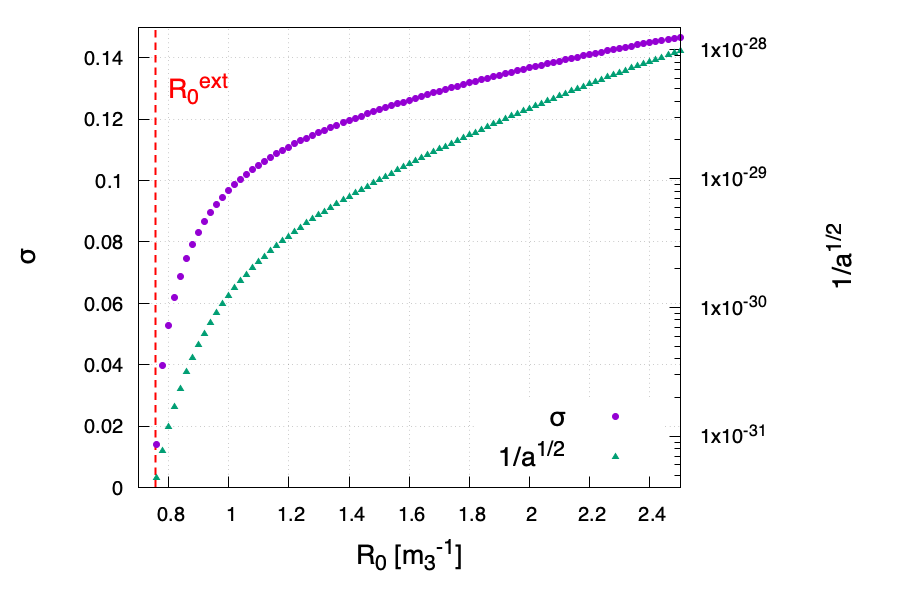}
    \caption{The parameter $\sigma$ and $1/\sqrt{a}$ as a function of the horizon size $R_0$.}
    \label{fig:sigma}
\end{figure}

In Fig.~\ref{fig:sigma}, we plot $\sigma$ and $1/\sqrt{a}$ as functions of $R_0$.
We observe that when $R_0$ approaches $R_0^\mathrm{ext}$, $\sigma$ approaches $0$ as expected. 
Similarly, when $R_0 = R_0^\mathrm{ext}$, the value of $a$ is consistent with our previous extremal result, which characterizes the deficit angle.

\begin{figure}
    \centering
    \includegraphics[width=0.5\textwidth]{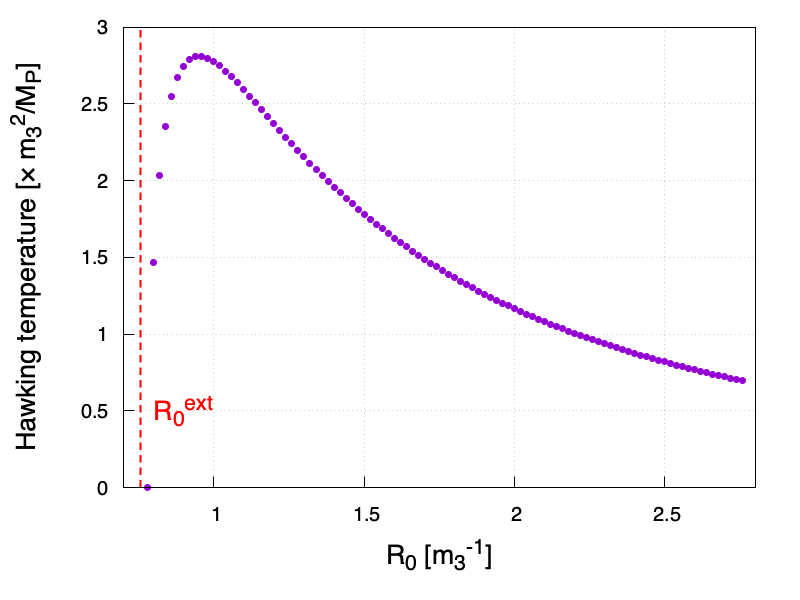}
    \caption{The Hawking temperature as a function of the horizon size $R_0$.}
    \label{fig:hawking_temperature}
\end{figure}

In the non-extremal case, the black string should have a non-zero temperature.
The Hawking temperature as a function of $R_0$ is plotted in Fig.~\ref{fig:hawking_temperature}. In this calculation, $A$ is normalized so that $A=1$ at a large $z$, instead of the way used in Sec.~\ref{sec:construction} ($\alpha = \beta = 1$). 
The behavior is very similar to that of the RN black hole. 

\begin{figure}
    \centering
    \includegraphics[width=1.05\linewidth]{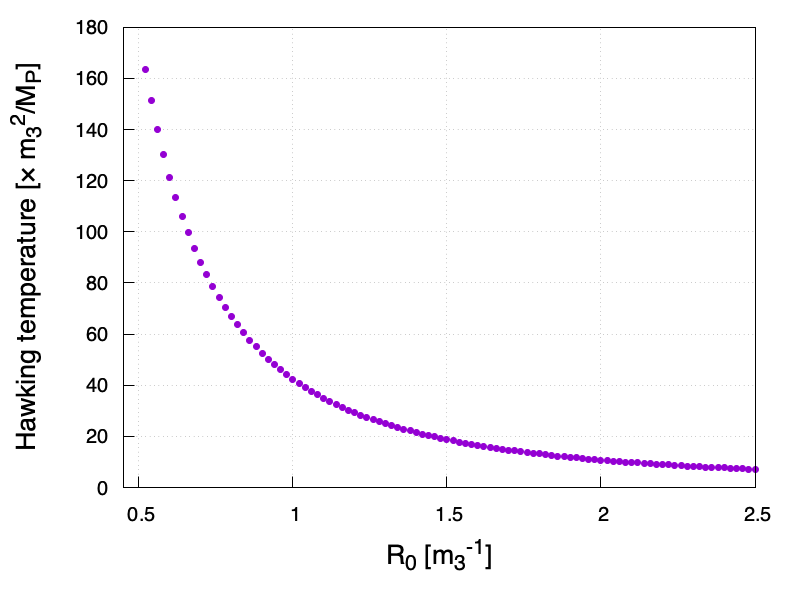}
    \caption{The Hawking temperature as a function of the horizon size $R_0$ in the case where the black string does not have $\mathbb{Z}_2$ charge.}
    \label{fig:hawking_temperature_zerocharge}
\end{figure}

It is valuable to see the behavior of the Hawking temperature when the string does not have $\mathbb{Z}_2$ charge. 
One can construct the black string solutions without $\mathbb{Z}_2$ charge by replacing the periodic boundary conditions for neutrinos in the Casimir energy with usual anti-periodic ones. 
In this case, there is no extremal solution as expected. 
Hence the Hawking temperature behaves as the function of horizon size as in the case of the Schwartzschild black hole, as shown in Fig.~\ref{fig:hawking_temperature_zerocharge}.
The black string may evaporate until the event horizon becomes Planckian in size.

\section{Discussion and Conclusion}\label{sec:discussion}
In this paper, we have constructed the black string solution.
There are several future directions to be explored.
For instance, the stability of the solution is important. The perturbative analysis based on the gauge invariant variable would be necessary to this end.
We have assumed the spin structure as a tangential structure of the spacetime. However, if the $B-L$ gauge field is introduced, it is more natural to consider the Spin$^c$ structure. 
In this case, the bordism group is $\Omega_1^{\text{Spin}^c}=0$. This implies the black string solution has an instability associated with the $B-L$ gauge field.

Another interesting direction is to construct the black solutions in string theories.
The same technology we have used in this paper could be applied to the string theory branes.\footnote{See Ref.~\cite{Fukuda:2024pvu} for the solutions of heterotic black $0$ and $4$-branes, and Ref.~\cite{Horowitz:1991cd} for the $6$-brane.}
We will report these issues in future publication~\cite{WIP}.
Furthermore, in the case of supersymmetric 7-branes~\cite{Greene:1989ya}, it involves an orbifold identification, which is viewed as an additional contribution to the deficit angle.
It is interesting to study a similar effect in our context.

It is also worth considering phenomenological and cosmological applications.
Particularly, it is quite non-trivial if the black string that we considered above can be produced in the Universe.
Since the size of the black string is very thick (even comparable to the Hubble horizon size for the temperature around $m_3$),
their dynamics after the production should be different from conventional cosmic strings~\cite{Vilenkin:2000jqa}.

\section*{Acknowledgments}
The authors would like to thank Tetsuya Shiromizu, Yuko Urakawa, and Hirotaka Yoshino for useful discussions.
The work of Y.H. is supported by the Deutsche Forschungsgemeinschaft under Germany's Excellence Strategy - EXC 2121 Quantum Universe - 390833306.
The work of Y.H. was supported by MEXT Leading Initiative for Excellent Young Researchers Grant No.JPMXS0320210099, JSPS KAKENHI Grants No.24H00976, 24K07035, and 24KF0167.


\bibliographystyle{apsrev4-1}
\bibliography{./reference}


\pagebreak
\widetext
\newpage
\begin{center}
\textbf{\large Supplemental Material for ``Black String in the Standard Model"}
\end{center}
\setcounter{figure}{0}
\setcounter{table}{0}
\setcounter{section}{0}
\setcounter{page}{1}

\medskip
This Supplemental Material includes the near-horizon geometry of the black string (Sec.~\ref{sec:interpolation}), derivation of the boundary condition at the horizon (Sec.~\ref{sec:boundary_condition}). 

\section{Extremal Black Hole and String as interpolating solutions}\label{sec:interpolation}

The near-horizon geometry of the extremal black solutions can be investigated by studying the effective potential for the radion field~\cite{Arkani-Hamed:2007ryu}.
Let us start from the case of the extremal Reissner-Nordstrom (RN) black hole.
By compactifying the 4d theory on $S^2$ with the magnetic flux, we obtain the effective 2d theory.
The effective 2d theory contains the radion field which parameterizes the size of $S^2$.
The radion potential consists of the flux and curvature terms, and the radion is stabilized at a minimum of the potential.
This minimum describes $AdS_2\times S^2$ spacetime, which is nothing but the near-horizon geometry of the extremal RN black hole.
In this sense, this black hole interpolates from Minkowski spacetime to $AdS_2 \times S^2$ vacuum.

\begin{figure}[h]
    \centering
    \includegraphics[width=0.4\textwidth]{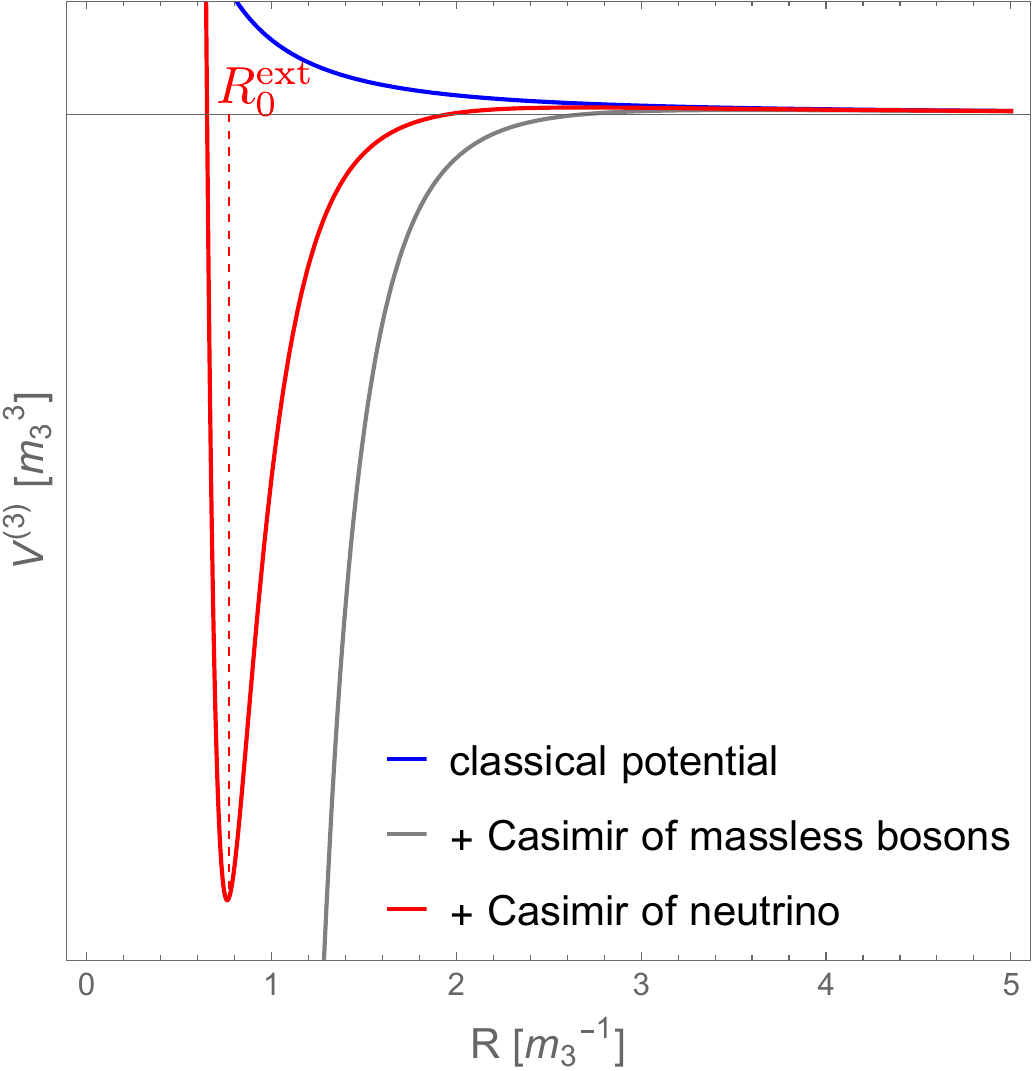}
    \caption{The effective potential for the radion, with no Casimir energy (blue line), the Casimir energy of massless particles (gray line), the Casimir energy of massless + neutrinos (red line).}
    \label{fig:radion_potential}
\end{figure}

The same is true for the extremal black string.
We can understand the near-horizon geometry of the extremal black string by compactifying the 4d theory on $S^1$.
We start from the action
\begin{align}
    S = \int d^4x \sqrt{-g^{(4)}}\left(\dfrac{1}{2}M_P^2\mathcal{R}^{(4)}-\Lambda^{(4)} - V_{\mathrm{Cas}}+ \ldots \right),\label{eq:action_of_sm}
\end{align}
where the superscript $(4)$ represents four-dimensional quantity, $\mathcal{R}^{(4)}$ is the Ricci scalar, $\Lambda^{(4)}$ is
the cosmological constant\footnote{We add the cosmological constant for the illustration though we assume zero cosmological constant in the main text.}, 
and $M_P$ is the reduced Planck mass.
We compactify the theory on $S^1$, and the effective 3D action becomes
\begin{align}
\begin{split}
    S = \int d^3x  \sqrt{-{g}^{(3)}} & \left[ \dfrac{1}{2}rM_P^2{\mathcal{R}}^{(3)} - rM_P^2 \left(\dfrac{\partial R}{R}\right)^2 \right. 
    - \left. \dfrac{rM_P^2}{8} \left( \dfrac{2\pi R}{r} \right)^2 R^2 F_{ij}F^{ij} \right.
    \left. 
    + V^{(3)}
    + \ldots \right],
\end{split}
\end{align}
where $R$ is the radion field, $F$ is the field strength of graviphoton, and $r$ is an arbitrary rescale factor with the length dimension. The potential for the radion is
\begin{align}
    V^{(3)} = -\dfrac{r^3 (\Lambda^{(4)} + V_{\mathrm{Cas}}(R))}{(2\pi R)^2}. \label{eq:radion_potential}
\end{align}
This potential has a minimum at the neutrino scale $R=R_0^\mathrm{ext}$ (see Fig.~\ref{fig:radion_potential}) when the neutrino mass is Majorana, and is the normal hierarchy with the lightest mass being $0$.
This minimum value is negative and describes $AdS_3\times S^1$ spacetime, which is nothing but the near-horizon geometry of the extremal black string~(\ref{eq:bc_extremal_horizon}, \ref{eq:bc_extremal_infty}).

\section{Boundary condition at horizon} \label{sec:boundary_condition}
For the non-extremal case, we first impose $A(0)=0$ at the horizon $z=0$ because the metric has a coordinate singularity at $z=0$. 
Moreover, as we are interested in the black string with the finite size, we impose $R(0) \neq 0$.
At the same time, $z=0$ should not be the curvature singularity.
This means that scalars quantities in the following,
\begin{align}
    &\frac{M_P^2}{m_3^4}\mathcal{R} = -2 \left(
        \frac{A' B'}{A B} + \frac{A' R'}{A R} + \frac{B' R'}{B R} \right.
        \left.  + \frac{A''}{A} + \frac{B''}{B} + \frac{R''}{R}
        \right)
        =-2 \left( \frac{A''}{A} + \frac{B''}{B} + \frac{R''}{R}-\tilde{V}_\mathrm{Cas} \right), \\
    &\frac{M_P^4}{m_3^8}\mathcal{R}_{\mu\nu}\mathcal{R}^{\mu\nu} =
        \left( \frac{A'}{A} \left( \frac{B'}{B} + \frac{R'}{R} \right) + \frac{A''}{A} \right)^2
        + \left( \frac{B'}{B} \left( \frac{A'}{A} + \frac{R'}{R} \right) + \frac{B''}{B} \right)^2 
        + \left( \frac{R'}{R} \left( \frac{A'}{A} + \frac{B'}{B} \right) + \frac{R''}{R} \right)^2,\\
    &\frac{M_P^4}{m_3^8}\mathcal{R}_{\mu\nu\rho\sigma}\mathcal{R}^{\mu\nu\rho\sigma} = 
    4 \left( 
        \frac{A'^2 B'^2}{A^2 B^2} + \frac{A'^2 R'^2}{A^2 R^2} + \frac{B'^2 R'^2}{B^2 R^2} \right.
        \left. + \frac{A''^2}{A^2} + \frac{B''^2}{B^2} + \frac{R''^2}{R^2} 
    \right),
\end{align}
do not diverge at $z=0$.
Since $\mathcal{R}_{\mu\nu}\mathcal{R}^{\mu\nu}$ and $\mathcal{R}_{\mu\nu\rho\sigma}\mathcal{R}^{\mu\nu\rho\sigma}$ are sums of semi-positive definite terms, each term should be finite.
Assuming that $A(z), B(z)$, and $R(z)$ are regular at $z=0$, we can expand them as a power series of $z$.
Then, we find
\begin{align}
    B^\prime(0) = 0, \ \ \ \ \ R^\prime(0) = 0, \label{eq:bc_at_horizon_2}
\end{align}
as well as $A^{\prime\prime}(0)=0$.

\end{document}